\documentclass[aip,jcp,amsmath,amssymb,reprint]{revtex4-1}

\usepackage[utf8]{inputenc}
\usepackage[T1]{fontenc}
\usepackage[english]{babel}
\usepackage[document]{ragged2e}

\usepackage{geometry} 
\usepackage{graphicx} 
\usepackage{color}
\usepackage{float}

\usepackage{booktabs} 
\usepackage{array} 
\usepackage{paralist} 
\usepackage{verbatim} 
\usepackage{subfig} 
\usepackage{hyperref}

\usepackage[justification=justified,format=plain]{caption}

\begin{document}

\title{Range-Separated Stochastic Resolution of Identity: Formulation and Application to Second Order Green's Function Theory} 

\author{Wenjie Dou}
\email{douw@berkeley.edu}
\thanks{These two authors contributed equally}
\affiliation{Department of Chemistry, University of California Berkeley, Berkeley California 94720, USA}

\author{Ming Chen}
\email{mingchen.chem@berkeley.edu}
\thanks{These two authors contributed equally}
\affiliation{Department of Chemistry, University of California Berkeley, Berkeley California 94720, USA}
\affiliation{Materials Sciences Division, Lawrence Berkeley National Laboratory, Berkeley, California 94720, USA}

\author{Tyler Y. Takeshita}
\email{tyler.takeshita@daimler.com}
\affiliation{Mercedes-Benz Research and Development North America, Sunnyvale, CA 94085}

\author{Roi Baer}
\email{roi.baer@huji.ac.il}
\affiliation{Fritz Haber Center for Molecular Dynamics, Institute of Chemistry, The Hebrew University of Jerusalem, Jerusalem 91904, Israel}

\author{Daniel Neuhauser}
\email{dxn@chem.ucla.edu}
\affiliation{Department of Chemistry and Biochemistry, University of California, Los Angeles, California 90095, USA}

\author{Eran Rabani}
\email{eran.rabani@berkeley.edu}
\affiliation{Department of Chemistry, University of California Berkeley, Berkeley California 94720, USA}
\affiliation{Materials Sciences Division, Lawrence Berkeley National Laboratory, Berkeley, California 94720, USA}
\affiliation{The Raymond and Beverly Sackler Center of Computational Molecular and Materials Science, Tel Aviv University, Tel Aviv 69978, Israel}

\begin{abstract}
We develop a range-separated stochastic resolution of identity approach for the $4$-index electron repulsion integrals, where the larger terms (above a predefined threshold) are treated using a deterministic resolution of identity and the remaining terms are treated using a stochastic resolution of identity. The approach is implemented within a second-order Greens function formalism with an improved $O(N^3)$ scaling with the size of the basis set,  $N$. Moreover, the range-separated approach greatly reduces the statistical error compared to the full stochastic version ({\it J. Chem. Phys.} {\bf 151}, 044144 (2019)), resulting in computational speedups of ground and excited state energies of nearly two orders of magnitude, as demonstrated for hydrogen dimer chains.
\end{abstract}

\maketitle

\section{Introduction}
\justify
Many-body perturbation theory (MBPT) based on Green's function (GF) approaches (e.g., the M{\o}ller-Plesset (MP) perturbation theory,\cite{PhysRev.46.618} the second order Green's function (GF2) approach,\cite{cederbaum1975one} the GW~\cite{PhysRev.139.A796} approximation) have been proven very useful in predicting ground state properties beyond the limitations of density functional theory (DFT) and the Hartree-Fock (HF) method, as well as in predicting quasi-particle and neutral excitation.  In these methods, correlations are treated systemically by expanding the self-energy (which contains the information of correlations) in the Coulomb~\cite{cederbaum1975one,Holleboom-1990} or screened Coulomb~\cite{Hybertsen1985,Rieger1999,RevModPhys.74.601} interactions.  MBPT has been applied to a variety of molecular and bulk systems in predicting, e.g. correlation energies, ionization potentials and electron affinities,\cite{dahlen2005variational,ohnishi2016explicitly,pavovsevic2017communication,Hybertsen1986,Rinke2005,Liao2011,Neaton2006,Tiago2006,Friedrich2006,Gruning2006,Shishkin2007,Rostgaard2010,PhysRevB.89.155417,Tamblyn2011,Marom2012,van2015gw} and excited states.\cite{PhysRevB.62.4927,PhysRevB.68.085310,Tiago2006,PhysRevB.91.235302,RevModPhys.74.601,Refaely-Abramson2011} Excluding several recent applications,\cite{PhysRevB.75.235102,PhysRevB.88.075105,PhysRevLett.113.076402,PhysRevB.85.081101,deslippe2012berkeleygw,foerster2011n,gonze2009abinit} MBPT has been limited to relatively small systems due to the steep computational scaling with the system size.
  
A particularly interesting implementation of MBPT, relevant to the applications reported below, is based on a second-order approximation to the electron self-energy,\cite{cederbaum1975one,Holleboom-1990,stefanucci2013nonequilibrium} which has received increasing attention in recent years. \cite{dahlen2005variational,phillips2014communication,ohnishi2016explicitly,pavovsevic2017communication} In contrast to the GW approximation,\cite{PhysRev.46.618} dynamical exchange correlations are included explicitly in the GF2 self-energy to second order in Coulomb interactions, providing  accurate ground state energies~\cite{kananenka2016efficient,rusakov2016self} and quasi-particle energies.\cite{dahlen2005self,welden2015ionization,ohnishi2016explicitly,pavovsevic2017communication} Although the results of recent studies are extremely promising, the GF2 approach suffers from a high computational cost ($O(N^5)$), limiting its application to relatively small system sizes.  

To overcome this limitation, two stochastic formulations were recently introduced to reduce the computational scaling.  Neuhauser {\em et al.}~\cite{neuhauser2017stochastic} developed a stochastic decomposition of the imaginary time GF to reduce the overall scaling of GF2 to  $O(N^3)$. Takeshita {\em et al.}~\cite{takeshita2019stochastic} and Dou {\em et al.}~\cite{dou2019stochastic} proposed an approach which builds upon the stochastic resolution of identity (SRI) for the electron repulsion integrals (ERIs)~\cite{Takeshita-2017stochastic}  to describe both ground and quasi-particle excited states. Similar to the deterministic resolution of identity (RI),\cite{Whitten-1973,Dunlap-1979-2,Dunlap-1979,Vahtras-1993,Feyereisen-1993} the SRI 
decouples the $4$-index ERIs; While the number of auxiliary basis increases with the system size for the RI, the number of stochastic orbitals in the SRI is independent of the system size, resulting in an overall $O(N^3)$ scaling.   However, the SRI technique comes at a cost of introducing a statistical error in the energy and nuclear forces,\cite{chen2019overlapped,arnon2020efficient,PhysRevB.97.115207,ge2013guided,Takeshita-2017stochastic} which can be controlled by increasing the number of stochastic realization, $N_s$. While the overall scaling of the stochastic formulations of GF2 is similar to DFT and HF, achieving chemical accuracy requires a large number of stochastic realization, resulting in increasingly longer computational time, even for small systems.\cite{takeshita2019stochastic,dou2019stochastic}

In this work, we develop a range-separated stochastic resolution of identity (RS-SRI) approach to decouple the $4$-index ERIs, where the short-range ERIs (larger values) are treated deterministically using the resolution of identity (RI)~\cite{Whitten-1973,Dunlap-1979-2,Dunlap-1979,Vahtras-1993,Feyereisen-1993} and the remaining terms are treated using the stochastic resolution of identity (SRI).\cite{Takeshita-2017stochastic}  The RS-SRI approach allows for a significant reduction of the statistical error without the need to increase the number of stochastic realization while maintaining the overall $O(N^3)$ scaling. We apply the RS-SRI technique to GF2 theory and demonstrate its ability to reduce the overall computational scaling from $O(N^5)$ to $O(N^3)$ as well as increase the sampling efficiency by nearly two orders of magnitude as compared to the SRI technique.  

\section{Range-separated stochastic resolution of identity}
\label{sec:theory}
Consider a generic many-body electronic Hamiltonian in the second-quantization representation:  
\begin{eqnarray} \label{eq:Hami}
\hat H = \sum_{ij} h_{ij}   \hat a^\dagger_i \hat a_j + \frac12 \sum_{ijkl}v_{ijkl}  \hat a^\dagger_i \hat a^\dagger_k  \hat a_l \hat a_j,
\end{eqnarray} 
where $\hat a^\dagger_i$ and $\hat a_i$ are the Fermionic creation and annihilation operators, respectively, for an electron in orbital $\chi_i ({\bf r})$. In the applications below, $\chi_i ({\bf r})$ is chosen to be an atomic orbital, but we do not use the locality of the basis to reduce the scaling nor do we introduce a cutoff to compute the ERIs (see Eq. \ref{eq:Smatrix}) or the overlap matrix (see Eq. \ref{eq:2e4c}). Therefore, the formalism and the resulting scaling reported below are general for any choice of basis.  The creation and annihilation operators obey the following anti-commutation relationship: 
\begin{eqnarray} \label{eq:Smatrix}
\left\{\hat a_i ,\hat a^\dagger_j \right\}=(\bold{S}^{-1})_{ij},
\end{eqnarray}
where $S_{ij} = \int \chi_i ({\bf r}) \chi_j ({\bf r}) d{\bf r}  $ is the matrix element of the overlap matrix $\bold{S}$.  In Eq.~(\ref{eq:Hami}), $h_{ij}$ is the matrix element of the one-body Hamiltonian and $v_{ijkl}$ is the $4$-index ERI ($v_{ijkl} \equiv (i  j | kl )$):
\begin{equation} 
	\label{eq:2e4c}
	v_{ijkl} = \iint d{\bf r}_1 d{\bf r}_2 \frac{\chi_i ({\bf r}_1)\chi_j ({\bf r}_1)\chi_k ({\bf r}_2)\chi_l ({\bf r}_2)}{\left|{\bf r}_1-{\bf r}_2\right|}.
\end{equation}
Describing correlations within a many-body perturbation technique beyond the mean-field approximation relies on contraction of $v_{ijkl}$ (or powers of $v_{ijkl}$), a task that becomes computationally intractable with increasing levels of accuracy. A common approach to reduce the computational complexity is based on the resolution of identity (RI), where the $4$-index ERIs in Eq.~(\ref{eq:2e4c}) are approximated by products of $3$-index ERIs and $2$-index ERIs:\cite{Whitten-1973,Dunlap-1979,Vahtras-1993,Feyereisen-1993}  
	\begin{equation}
	 \label{eq:abgd-stoch}
	\begin{split}
		v_{ijkl} &\approx \sum_{AB}^{N_{\rm aux}} (i j |A)  V^{-1}_{AB} (B| kl ).
	\end{split}
	\end{equation}
Here, $\chi_A({\bf r})$ and $\chi_B({\bf r})$ are auxiliary orbitals, and $(i j |A)$ and $V_{AB}$ are $3$-index and $2$-index ERIs respectively,  
\begin{equation}
		(i j |A) = \iint  d{\bf r}_1 d{\bf r}_2 \frac{\chi_i ({\bf r}_1)\chi_j ({\bf r}_1)\chi_A({\bf r}_2)}{\left|{\bf r}_1-{\bf r}_2\right|}
	\end{equation}
	\begin{equation}
		V_{AB} = \iint  d{\bf r}_1 d{\bf r}_2 \frac{\chi_A({\bf r}_1) \chi_B({\bf r}_2)}{\left|{\bf r}_1-{\bf r}_2\right|}.
	\end{equation}
For convenience, we define a new set of $3$-index ERIs $K_{ij}^Q$
\begin{eqnarray} \label{eq:K_ijQ}
K_{ij}^Q = \sum_A^{N_{\rm aux}} (ij|A) V_{AQ}^{-\frac12}
\end{eqnarray}
such that the 4-index ERI can be expressed in terms of $3$-index ERIs only: 
\begin{eqnarray} \label{eq:vKQKQ}
v_{ijkl} & =  \sum_{Q}^{N_{\rm aux}} K_{ij}^Q K_{kl}^Q.
\end{eqnarray}
The advantage of the above decomposition is that the resolution of identity reduces the number of $2$-body ERIs from O($N^4$) to  O($N^2 N_{\rm aux}$), where $N$ is the size of the atomic basis and $N_{\rm aux}$ is the size of the auxiliary basis. However, since  $N_{\rm aux}$ increases nearly linearly with the size of the atomic basis $N$ and since the calculation of $K_{ij}^Q$ scales as O($N^4$), the approach does not always reduce the computational scaling of the correlation energy for, e.g., MP2 and GF2.\cite{Takeshita-2017stochastic,takeshita2019stochastic,dou2019stochastic}

Recently, we have introduced a stochastic version of the resolution of identity, which provides a framework to reduce the scaling for contraction within many-body perturbation techniques at the account of introducing a controlled statistical error in the calculated observables (e.g. the forces on the nuclei, the energy per electron). The balance between accuracy and efficiency is controlled by the number of stochastic realizations ($N_s$) according to the central limit theorem. The stochastic RI approach utilizes the same set of $3$-index ERIs $(ij|A)$ while circumventing the need to directly compute $K_{ij}^Q$ by introducing a set of $N_s$ \textit{stochastic orbitals}, $\{ \theta^\xi \}$, $\xi = 1, 2, \cdots, N_s$. The stochastic orbitals are defined as arrays of length $N_{\rm aux}$ with randomly selected elements 1 or -1, i.e. $\theta_A^\xi = \pm1$. Defining 
\begin{eqnarray} \label{eq:4}
         {R_{ i j }^{\xi} } &=& \sum_{AQ}^{N_{\rm aux}} (ij|A) V_{AQ}^{-\frac12}\theta_Q^{\xi} \nonumber \\& =& \sum_{A}^{N_{\rm aux}} (ij|A) \sum_{Q}^{N_{\rm aux}} V_{AQ}^{-\frac12}\theta_Q^{\xi} 
\end{eqnarray}
the expression for $v_{ijkl}$ can be reduced to:
\begin{equation}
	\label{eq:abgd_l}
	v_{ijkl} \approx \frac{1}{N_s} \sum_\xi  R_{ij}^{\xi} R_{kl}^{\xi} \equiv \left< R_{ij} R_{kl}  \right>_\theta,
	\end{equation}
where $\left< \cdots \right>_\theta$ implies a statistical average over the stochastic orbitals, $\{\theta\}$. The overall computational scaling of the $R_{ij}^{\xi}$ matrices is $O(N_s N^3)$, but $N_s$ is found to be independent of the system size for different applications.\cite{Takeshita-2017stochastic,takeshita2019stochastic,dou2019stochastic,chen2019overlapped,PhysRevB.91.235302,gao2015sublinear,PhysRevLett.113.076402,gao2015sublinear}  The SRI technique has been successfully used to reduce the scaling of the correlation energy within MP2 and GF2 theories, from $O(N^5)$ to $O(N^3)$. 

The above approach has been implemented for simple molecules and for hydrogen chains of different length in order to assess its accuracy for large systems.\cite{Takeshita-2017stochastic,takeshita2019stochastic,dou2019stochastic}  To converge the results to chemical accuracy required a rather large number of stochastic orbitals ($N_s \approx 1000$), which limits the application of the SRI technique to relatively small systems (due to the large "prefactor"), with $N\rightarrow 1000$, still exceedingly larger than the deterministic approach.\cite{Takeshita-2017stochastic,takeshita2019stochastic,dou2019stochastic} In order to reduce the number of stochastic orbitals and to allow for a smaller statistical error, we first sort the ERIs $(ij|A)$ according to their magnitude and keep only those that are larger than a threshold:
\begin{widetext}
\begin{eqnarray} \label{eq:V_ijA}
(ij|A)^{L} = 
\begin{cases}
(ij|A)     & \quad \text{if } |(ij|A)| \ge \frac{\epsilon^\prime}{N} {\{|(ij|A)|\}}^{\rm max}_j \\
0  & \quad \text{otherwise}.
\end{cases}
\end{eqnarray}
\end{widetext}
Here, ${\{|(ij|A)|\}}^{\rm max}_j$ is the maximal value of $|(ij|A)|$ for each $j$ and $\epsilon^\prime$ is a predefined parameter. The superscript $L$ (or $S$ denotes large (or small) values. By setting the cutoff threshold to depend on $\frac{\epsilon^\prime}{N}$, the number of nonzero elements in $(ij|A)^{L}$ for each $j$ scales as $O(N)$ (rather than $O(N^2)$ if no threshold is used or $O(1)$ if fixed threshold $\epsilon^\prime$ is used). This implies that the total non-vanishing elements in $(ij|A)^{L}$ scales as O($N^2$). We then define $[K_{ij}^Q]^{L}$ as 
\begin{eqnarray} \label{eq:K_ijQ}
[K_{ij}^Q]^{L} = \sum_A^{N_{\rm aux}} (ij|A)^{L} V_{AQ}^{-\frac12}
\end{eqnarray}
and keep only the terms that are larger than a predefined threshold, namely, we set  ${[K_{ij}^Q]}^{L} =0$ for values below the threshold according to:
\begin{eqnarray} \label{eq:K_L}
{[K_{ij}^Q]}^{L} = 
\begin{cases}
[K_{ij}^Q]^{L}     & \quad \text{if } |[K_{ij}^Q]^{L}| \ge {\epsilon} {\{|[K_{ij}^Q]^{L}|\}}^{\rm max} \\
0  & \quad \text{otherwise} ,
\end{cases}
\end{eqnarray}
The calculation of $[K_{ij}^Q]^{L} $ using the above procedure scales as $O(N^3)$.  We proceed by defining: 
\begin{eqnarray}
         {[R_{ i j }^{\xi} ]}^{L} = \sum_Q^{N_{\rm aux}}  {[K_{ij}^Q]}^{L} \theta_Q^{\xi} \\
         {[R_{ i j }^{\xi} ]}^{S} =  {R_{ i j }^{\xi} } -  {[R_{ i j }^{\xi} ]}^{L},
\end{eqnarray}
where ${R_{ i j }^{\xi} }$ is defined above in Eq.~(\ref{eq:4}) and the computational scaling for both terms, ${[R_{ i j }^{\xi} ]}^{L} $ and ${[R_{ i j }^{\xi} ]}^{S} $, is $O(N^3)$. Using these definitions, the $4$-index tensor $v_{ijkl}$ can be rewritten as:
\begin{eqnarray} \label{eq:RS-SRI}
v_{ijkl} & =&  \sum_Q^{N_{\rm aux}} {[K_{ij}^Q]}^L {[K_{kl}^Q]}^L +  \left< R_{ij}^L R_{kl}^S  \right>_\theta \nonumber \\& +&  \left< R_{ij}^S R_{kl}^L  \right>_\theta +  \left< R_{ij}^S R_{kl}^S  \right>_\theta
\end{eqnarray}
Eq.~(\ref{eq:RS-SRI}) is referred to as range-separated stochastic resolution of identity (RS-SRI). The RS-SRI reduces to the SRI for $\epsilon=1$ and to the deterministic RI for $\epsilon=0$. This suggest that $\epsilon$ can be used as a control parameter balancing the computational efficiency and statistical errors. For optimal choices of $\epsilon$, the contribution of $\sum_Q^{N_{\rm aux}} {[K_{ij}^Q]}^L {[K_{kl}^Q]}^L$ in Eq.~(\ref{eq:RS-SRI}) must be larger than the other terms. 

\section{Application to Second Order Green's Function}
\label{sec:GF2}
We now apply the above formalism to the second order Matsubara Green's function (GF2) theory.\cite{takeshita2019stochastic,dou2019stochastic,neuhauser2017stochastic} The main entity in the GF2 theory is the Matsubara single-particle, finite temperature, Green's function given by (we set $\hbar=1$ unless otherwise stated):
\begin{eqnarray}
G_{ij} (\tau )  = - \langle T_c \hat a_i
(\tau) \hat a^\dagger_j  \rangle,
\label{eq:Gij}
\end{eqnarray}
where  $\hat a_i$ and $\hat a_j^\dagger$ are defined above in Sec. \ref{sec:theory}, $T_c$ is a time ordering operator, and $\tau$ is an imaginary time point along $\tau \in (0, -\beta)$. In the above, we have used the Heisenberg picture for the operators: $\hat a_i (\tau) =
e^{ (\hat H -\mu \hat N) \tau } \hat a_i e^{- (\hat H -\mu \hat N) \tau }$, where $\hat N = \sum_{ij} S_{ij} \hat a^\dagger_i \hat a_j $ is the number operator and $\hat{H}$ is the many-body Hamiltonian defined in Eq.~(\ref{eq:Hami}). The average is taken with respect to the grand canonical partition function: $\langle \cdots \rangle = Z^{-1} \mbox{Tr}
\left[(\cdots) e^{-\beta (\hat H - \mu \hat N) } \right]$, where $Z =
\mbox{Tr}\left[e^{-\beta (\hat H - \mu \hat N)}\right]$ is the normalization factor, $\beta=1/{k_{\rm B}T}$ is the inverse temperature, and $\mu$ is the chemical potential. 

The Matsubara GF obeys the following Dyson equation: 
\begin{eqnarray} \label{eq:EOM}
-\bold{S} \partial_\tau \bold{G}(\tau) &=& \delta (\tau) + (\bold{F} - \mu \bold{S} )\bold{G}(\tau) \nonumber \\ &+& \int_0^\beta d\tau_1 \bold{\Sigma} (\tau - \tau_1) \bold{G} (\tau_1)  
\end{eqnarray}
where $\bold{F}$ is the Fock matrix given by: 
\begin{eqnarray} \label{eq:F}
F_{ij} = h_{ij} - 2 \sum_{kl} G_{kl} (\beta^-) (v_{ijkl} - \frac12 v_{ilkj}){{~}}{{~}}{{~}}
\end{eqnarray}
and $\bold{\Sigma}$ is the self-energy. In the second-order Born approximation, the self-energy (in the closed shell case) is given by:
\begin{eqnarray}
\label{eq:sigma}
\Sigma_{ij} (\tau) &=& \sum_{klmnpq} v_{imqk}(2v_{lpnj} - v_{nplj}) \nonumber \\  &\times &G_{kl}(\tau)G_{mn}(\tau)G_{pq} (\beta-\tau) .
\end{eqnarray}
The above form scales as $O(N^5)$ using the appropriate contraction. 

The Matsubara Green's function for the Fermionic systems obeys the following anti-symmetric relationship: $\bold{G}(\tau) = - \bold{G} (\tau+ \beta)$. The anti-symmetry feature allows for a Fourier representation of $\bold{G}(\tau)$ in imaginary frequency: 
\begin{eqnarray} \label{eq:GFT}
\mathbf{\tilde
            G}(i\omega_n) = \int_0^\beta e^{i\omega_n \tau} \mathbf{G}(\tau) .
\end{eqnarray}  
Here, $i\omega_n = i (2n+1) \frac{\pi} \beta $ are the Matsubara frequencies and the inverse Fourier transform is defined by: 
\begin{eqnarray} \label{eq:Ginverse}
\mathbf{G}(\tau) = \frac{1}{\beta}\sum_n e^{-i\omega_n \tau}\mathbf{\tilde G}(i\omega_n).
\end{eqnarray}            
The Dyson equation (cf., Eq.~(\ref{eq:EOM})) can then be solved in the frequency domain: 
\begin{equation}
\label{eq:dyson}
\mathbf{\tilde G}(i\omega_n) = \frac{1}{[\mathbf{\tilde G}_0(i\omega_n)]^{-1}- \mathbf{\tilde \Sigma}(i\omega_n)},
\end{equation}
where $\mathbf{\tilde \Sigma}(i\omega_n)$ is the Fourier transform of the self-energy (Eq.~(\ref{eq:sigma})) and $\mathbf{\tilde G_0}(i\omega_n)$  is the non-interacting GF: 
\begin{eqnarray} 
\label{eq:G0}
\mathbf{\tilde G_0}(i\omega_n)= [(\mu + i\omega_n)\mathbf{S} - \mathbf{F}]^{-1}\\ \nonumber
\end{eqnarray}
Since the self-energy $\mathbf{\tilde \Sigma}(i\omega_n)$ depends on $\mathbf{\tilde G}(i\omega_n)$ itself, Eq.~(\ref{eq:dyson}) as well as Eq.~(\ref{eq:sigma}) must be solved self-consistently. This is done by first performing a Hartree-Fock calculation to obtain the overlap matrix $\mathbf{S}$, the Fock
        matrix $\mathbf{F}$ and the chemical potential $\mu$. The Fock matrix can then be used for constructing the non-interacting GF (cf., Eq.~(\ref{eq:G0}) which serves as our initial guess of $\mathbf{\tilde G}(i\omega_n) = \mathbf{\tilde G_0}(i\omega_n)$.  The next step involves the calculation of the self-energy, which is preformed in the imaginary time domain (Eq.~(\ref{eq:sigma})). The self-energy is then used to update the GF in Eq.~(\ref{eq:dyson}) and the latter is used to update the Fock matrix in Eq.~(\ref{eq:F}). It is often necessary to conserve the number of particles $N_e = -2 \sum_{ij} G_{ij} (\tau = \beta^-)  S_{ij}$. This can be achieved by tuning the chemical potential $\mu$. 

The computational bottleneck in GF2 is the calculation is the self-energy, which scales formally as $O(N^5)$. Using the RS-SRI representation for $v_{ijkl}$ given by Eq.~(\ref{eq:RS-SRI}), the self-energy can be written as:
\begin{widetext}
\begin{eqnarray} \label{eq:sigma_2}
\Sigma_{ij} (\tau) &=&
\sum_{klmnpq}G_{kl}(\tau)G_{mn}(\tau)G_{pq}(\beta-\tau)  \nonumber \\
&\times& \left(\sum_Q^{N_{\rm aux}} {[K_{im}^Q]}^L {[K_{qk}^Q]}^L +  \left< R_{im}^L R_{qk}^S  \right>_\theta +  \left< R_{im}^S R_{qk}^L  \right>_\theta +  \left< R_{im}^S R_{qk}^S  \right>_\theta \right)  \nonumber \\
& \times& \Bigg[ 2 \left( \sum_Q^{N_{\rm aux}} {[K_{lp}^Q]}^L {[K_{nj}^Q]}^L +  \left< R_{lp}^L R_{nj}^S  \right>_{\theta'} +  \left< R_{lp}^S R_{nj}^L  \right>_{\theta'} +  \left< R_{lp}^S R_{nj}^S  \right>_{\theta'}  \right) \nonumber \\
&-& \left( \sum_Q^{N_{\rm aux}} {[K_{np}^Q]}^L {[K_{lj}^Q]}^L +  \left< R_{np}^L R_{lj}^S  \right>_{\theta'} +  \left< R_{np}^S R_{lj}^L  \right>_{\theta'} +  \left< R_{np}^S R_{lj}^S  \right>_{\theta'}  \right) \Bigg]
\end{eqnarray}
\end{widetext}

In the following section we apply the RS-SRI to a series of hydrogen chain molecules and compare the results to deterministic RI as well as to SRI.  We find in practice that the RS-SRI scales even better than the upper theoretical limit of $O(N^3$) and at the same time reduces the statistical  error by about an order of magnitude as shown below.

\section{Results and Discussion}
\label{sec:results}
\justify
In this section, we assess the performance of the RS-SRI-GF2 approach and compare the results to deterministic and SRI-GF2 for hydrogen dimer chains $H_{N_H}$ of length $N_H$. The distance between strongly bonded hydrogen atoms was set to $0.74$~{\AA} and the distance between  weakly bonded  hydrogen atoms was set to $1.26$~{\AA}. For each hydrogen, we used the STO-3G basis and the CC-pVDZ-RI fitting basis for the resolution of identity in evaluating the self-energy as well as CC-pVDZ-JKFIT fitting basis in evaluating the Fock matrix in Eq.~(\ref{eq:F}). The inverse temperature used for the calculation of the GFs was set to $\beta = 50$~inverse Hartree, sufficient to converge the results due to the large quasi-particle gap. We used the approach developed in Ref.~\onlinecite{neuhauser2017stochastic} to perform the discrete Fourier transform with  $20,000$ Matsubara frequencies and $300$ imaginary-time points. We also have set $\epsilon' = 0.02$ in Eq.~(\ref{eq:V_ijA}) and $\epsilon = 0.1$ in Eq.~(\ref{eq:K_L}) as our thresholds for RS-SRI calculation below. 

\begin{figure}[t] 
   \centering{}
   \justify
   \includegraphics[width=8cm]{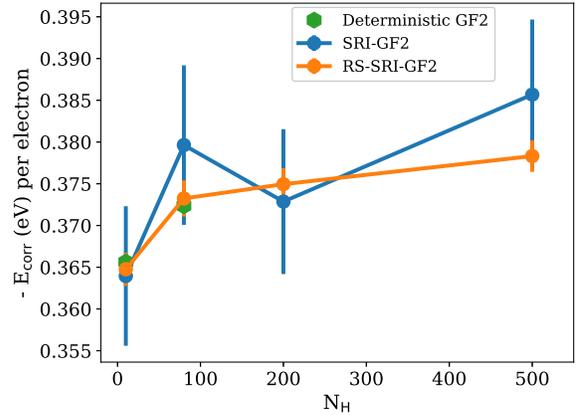} 
   \caption{Correlation energy per electron (cf., Eq.~(\ref{eq:corr})) for a series of Hydrogen dimer chains of different length ($N_H$ is the number of hydrogen atoms). The error bar is estimated by the standard deviation of the mean values, $\frac{\sigma}{\sqrt{N_{\rm samp}}}$. We have used $N_s = 800$ stochastic orbitals for both RS-SRI-GF2 and SRI-GF2 calculations. Note that both stochastic approaches agree with the deterministic approach (calculated only for the smaller system sizes) within the statistical error.\cite{RS_SRI_notes2}}
   \label{fig:corr}
\end{figure}

In Fig. \ref{fig:corr}, we plot the correlation energy per electron, defined as~\cite{neuhauser2017stochastic}
\begin{eqnarray}
E_{\rm corr} = \frac{1}{N_e}\int_0^\beta d\tau~\mathrm{Tr}~(\mathbf \Sigma(\tau) \mathbf G(\beta - \tau) )
\label{eq:corr}
\end{eqnarray}
for a series of Hydrogen dimer chains. We compare the results obtained using the RS-SRI-GF2 with SRI-GF2 and for small systems, with deterministic calculations. We find, as expected, that the correlation energy per electron is roughly independent of the length of the chain. Furthermore, both RS-SRI-GF2 with SRI-GF2 agree with the deterministic results within their statistical error. However, the statistical error for the same number of stochastic orbitals  ($N_s$) is significantly smaller (by nearly an order of magnitude) for RS-SRI-GF2 compared to SRI-GF2 for the entire range of systems sizes. The error bar was estimated as the standard deviation of the mean values, $\frac{\sigma}{\sqrt{N_{\rm samp}}}$, where $N_{\rm samp}=10$ was the number of samples used to estimate the statistical fluctuations. 

\begin{figure}[t] 
   \centering
   \includegraphics[width=8cm]{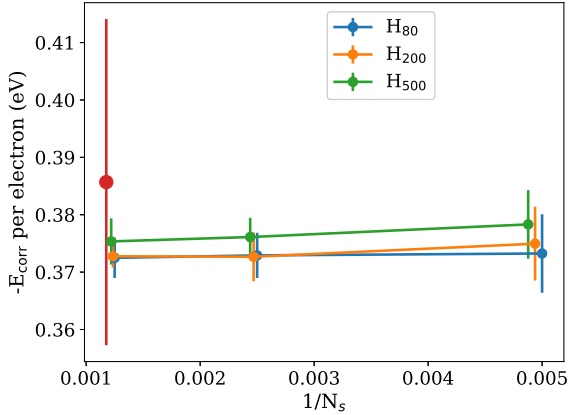} 
   \caption{ The correlation energy per electron as a function of $1/N_s$ for H$_{80}$, H$_{200}$, H$_{500}$ obtained using the RS-SRI-GF2. For $N_s=800$ we also show the result for $N_H=500$ using the SRI-GF2 approach (red symbol).  Note that for clarity we have shifted slightly the values of the $x$ axis for the difrerent system sizes.}
   \label{fig:error}
\end{figure}

In Fig. \ref{fig:error}, we plot the correlation energy per electron as a function of the inverse of the number of stochastic orbitals ($\frac{1}{N_s}$) for H$_{80}$, H$_{200}$, H$_{500}$. We find that the statistical  fluctuations decrease as $\frac{1}{\sqrt{N_s}}$, indicated by the decrease in the magnitude of the error bars. For $N_s=800$ we compare the RS-SRI-GF2 with the SRI-GF2 (red symbol, Fig.~\ref{fig:error}) for H$_{500}$. Clearly, the statistical noise is much larger (by about a factor of $10$) compared to the RS-SRI-GF2 result (green symbols).  We also find that the statistical fluctuations in the correlation energy per electron are independent of the system size. However, for the largest system studied, e.g. H$_{500}$, we observe a bias, where the correlation energy per electron decreases linearly with $\frac{1}{N_s}$. In Ref.~\onlinecite{neuhauser2017stochastic} the authors also report on the existence of bias. This results from the self-consistent treatment, but in comparison to previous work, the current bias is negligibly small, well within the statistical errors and thus, its existence is questionable.

\begin{figure}[t] 
   \centering
   \includegraphics[width=7.75cm]{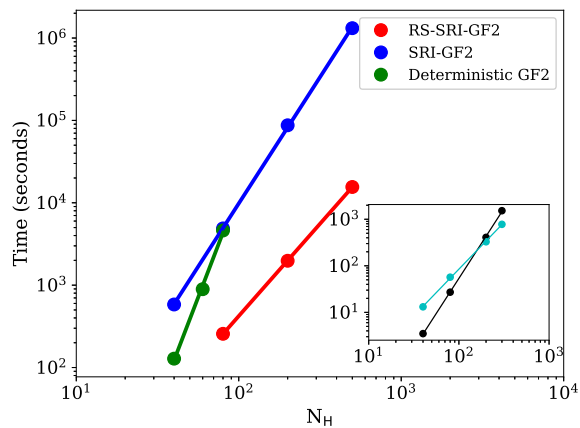} 
   \caption{Computational wall time of the different GF2 approaches (deterministic GF2, RS-SRI-GF2, and SRI-GF2) as a function of $N_H$.  The inset shows the scaling of computing $[K_{ij}^Q]^{L}$ (black symbols) as well as the scaling of the deterministic portion of the self-energy (terms that only involve $[K_{ij}^Q]^{L}$ but not $R_{ij}^L$ or $R_{ij}^S$, cyan symbols). }
   \label{fig:RS_timing}
\end{figure}

In Fig. \ref{fig:RS_timing}, we plot the computational wall time of the different GF2 approaches (deterministic GF2, RS-SRI-GF2, and SRI-GF2) as a function of the length of the hydrogen atom chain, $N_H$. All calculations are performed on a single node with the 32-core Intel-Xeon processor
E5-2698 v3 (“Haswell”) at 2.3 GHz. The deterministic GF2 scales as $O(N^{5.1})$, the SRI-GF2 scales as $O(N^{3.1})$, and the current approach, for the same level of accuracy as in the SRI-GF2, scales as $O(N^{2.2})$, slightly better than theoretical limit of $O(N^{3})$. Note that the RS-SRI-GF2 approach has a much smaller total wall time compared to the other approaches, across the entire system range studied. As additional checks, the inset of Fig.~\ref{fig:RS_timing} shows the scaling of computing $[K_{ij}^Q]^{L}$ as well as the scaling of the deterministic portion of the self-energy (terms that only involve $[K_{ij}^Q]^{L}$ but not $R_{ij}^L$ or $R_{ij}^S$). The former scales as $O(N^3)$ and the latter is found to scale as $O(N^2)$.

\section{Conclusions}
\label{sec:conclusions} 
We have developed a range-separated stochastic resolution of identity approach to decouple the $4$-index electron repulsion integrals and implemented the approach within the second order Green's function formalism. The RS-SRI technique can be viewed as a hybridization of the RI and SRI techniques, leveraging from both the accuracy of the RI and the reduced computational complexity of the SRI approaches. Results calculated for hydrogen dimer chains of varying length show an improved scaling of $O(N^{2.2}$) with the size of the basis, $N$. In comparison to our previous fully stochastic approach, the RS-SRI-GF2 approach reduces significantly the statistical error, resulting in computational wall times that are nearly two orders of magnitude shorter compared to the SRI-GF2. While we focused in this work on the specific implementation of the RS-SRI, the approach lends itself to higher-order approximations to the self-energy and for going beyond ground state properties. Future work should assess the performance of this RS-SRI technique for a wider range of geometries as well as its applicability to calculation of excited state properties. 

\begin{acknowledgements}
D.N. and E.R. are grateful for
support by the Center for Computational Study of Excited State
Phenomena in Energy Materials (C2SEPEM) at the Lawrence Berkeley
National Laboratory, which is funded by the U.S. Department of Energy,
Office of Science, Basic energy Sciences, Materials Sciences and
Engineering Division under Contract No. DEAC02-05CH11231 as part of
the Computational Materials Sciences Program. R.B. is grateful for
support by Binational US-Israel Science Foundation grant BSF-2020602. Resources of the
National Energy Research Scientific Computing Center (NERSC), a U.S.
Department of Energy Office of Science User Facility operated under
Contract No. DE-AC02-05CH11231, are greatly acknowledged.
\end{acknowledgements}

\section*{DATA AVAILABLITY}
The data that support the findings of this study are available from the corresponding author upon reasonable request.



%


\begin{thebibliography}{55}
\bibitem [{RS_()}]{RS_SRI_notes2}%
  \BibitemOpen
  \href@noop {} {}\bibinfo {note} {Due to the steep scaling of $O(N^5)$, we
  were reluctant to compute the correlation energies deterministically for
  $N>100$.}\BibitemShut {Stop}%
\end{thebibliography}
\end{document}